**Complexity**

# A session-based song recommendation approach involving user characterization along the play power-law distribution

Diego Sánchez-Moreno, Vivian F. López Batista, M. Dolores Muñoz Vicente, Ana B. Gil González and María N. Moreno-García

Department of Computing and Automation, University of Salamanca, Salamanca,37008, Spain.

Correspondence should be addressed to María N. Moreno; mmg@usal.es

## Abstract

In recent years, streaming music platforms have become very popular mainly due to the huge number of songs these systems make available to users. This enormous availability means that recommendation mechanisms that help users to select the music they like need to be incorporated. However, developing reliable recommender systems in the music field involves dealing with many problems, some of which are generic and widely studied in the literature, while others are specific to this application domain and are therefore less well-known. This work is focused on two important issues that have not received much attention: managing gray-sheep users and obtaining implicit ratings. The first one is usually addressed by resorting to content information that is often difficult to obtain. The other drawback is related to the sparsity problem that arises when there are obstacles to gather explicit ratings. In this work, the referred shortcomings are addressed by means of a recommendation approach based on the users' streaming sessions. The method is aimed at managing the well-known power-law probability distribution representing the listening behavior of users. This proposal improves the recommendation reliability of collaborative filtering methods while reducing the complexity of the procedures used so far to deal with the gray-sheep problem.

## 1. Introduction

In the digital era, where e-commerce and digital content distribution are so extended, recommender systems have become indispensable tools to help users to find the information, products or services they are interested in. These systems are especially useful in the area of music streaming services, given the large volume of content they make available to listeners. Most streaming platforms have advanced filtering mechanisms and even music recommender systems. However, user satisfaction data indicates that their reliability is not very high [17], [19]. This may be due to numerous problems with the recommendation methods that occur irrespective of the application domain, as well as those specific to the field of music.

Collaborative filtering (CF) is the most extended recommendation approach and one of the most reliable. Its main characteristic is the use of ratings given by users to items to be recommended. The ratings are stored in a U x I matrix, where U is the number of users and I the number of items in the system. The GroupLens research system for Usenet news [38] was the first recommender system using CF and Ringo [44] was one of the first and most popular music recommender systems based on CF.

There are two categories of CF methods, user-based and item-based. In the first, the active user receives the recommendations of items that have been positively rated by other users with similar tastes to him, that is, his/her nearest neighbors. These users have rated items in common with the active user in a similar way. Neighborhood can be computed by means of different similarity metrics. The most widely used ones are the Pearson correlation coefficient and cosine similarity [3]. Since the nearest neighbors are searched at recommendation time, user-based CF methods are also called memory-based. One of their main problems is scalability, which causes an exponential increment of the user response time as the number of users and the number of products in the system increase. In order to avoid this problem, item-based CF was proposed [44]. In this approach, rating-based similarities between items are computed before recommendation time, and then the active user receives suggestions of items similar to those he/she previously rated positively. This can be done since it is expected that new ratings given to items in large databases do not significantly change the similarity between them, especially for much rated items. This type of methods are also called as model-based methods since they make use of a model induced before the active user accesses the system. However, recommendations provided by item-based methods usually have less quality than those provided by user-based approaches. Therefore, they are indicated to be applied in large-scale systems where scalability is a serious problem. That is the case of Amazon, a very popular system where item-based models have been used [25].

The need for the explicit expression of the user's personal preferences for items in the form of ratings is the cause of the other major drawback of CF: the sparsity problem, which arises when the rating matrix contains a large number of null elements. This means that the number of ratings obtained from the users is fewer than the number of ratings needed for prediction [28]. Matrix factorization approaches can be used to deal with this problem, but these methods have some disadvantages, such as the cost of building the models and the loss of information resulting from the dimensionality reduction, and these are not always compensated with a significant improvement of results [43]. Thus, in many cases it is more effective to resort to implicit ratings that can be obtained from the time that users spend examining the items or from other data stored in log files,  although in this case, it must be assumed that preferences derived from this information are usually not as reliable as the explicit ones.

Content-based methods are alternative approaches to CF that base recommendations on the similarity between items, as item-based techniques. Nevertheless, they do not need rating data since they make use of other features of the items for computing the similarity. These methods can be applied to address two well-known shortcomings of both user-based and item-based CF: early-rater (first-rater) and cold-start. The first drawback is observed when new products are introduced into the system. These items have never been rated; therefore, they cannot be recommended. The cold-start problem affects new users, who cannot receive recommendations because they have no or few evaluations about products. In these circumstances, item content is used to make recommendations of items similar to those that the user likes. Content-based methods are also used to address the gray-sheep problem suffered by users with unusual tastes, for whom it is very difficult to find neighbors [8].

Other proposals to deal with the above-mentioned shortcomings of recommendation methods seek to build on the strengths of every category and avoid their weaknesses by means of hybrid approaches. This class of methods, which is currently the most extended, involves the combination of either different types of CF or CF with content-based techniques, among others. [49].

Music recommender systems have other additional limitations that are specific to this application domain. On the one hand, explicit ratings are usually not available in the streaming platforms, so ratings are obtained from implicit feedback. This is the main difference between music and other domains: while items such as books and other products can be evaluated from their purchase records, musical items in the streaming platforms cannot be evaluated in that way because they are not purchased individually. Another difference is the way these items are consumed. While a book or a movie is generally read or watched by a given user once, a song is usually listened to many times. This quality can be used to derive ratings from the number of times that users play a given song or artist, but this is not a trivial task, given its characteristic frequency distribution. The frequency of plays of musical items (artists or songs) adopts a power-law distribution, since high frequencies of plays are concentrated in very few items, while the remaining ones are part of the long tail of the curve [6]. Simple frequency functions usually used to transform plays into ratings are not suitable in this case.

The complexity of recommender systems has been increasing in recent years as they have evolved from initial CF or content-based systems to current systems that mostly use hybrid methods. In the latter, the level of complexity of the procedures and the information to be processed is much higher. While the basic collaborative filtering methods use simple algorithms such as k-nearest neighbors or matrix factorization, the hybrid methods combine these techniques with more complex machine learning algorithms. In addition, current systems do not only use explicit preference information, but are able to infer that knowledge from user behavior and manage other user an item attributes. This usually requires, unlike traditional systems, collecting and processing not only static but also dynamic information, which entails greater difficulty.

The work presented in this paper deals with that complexity while addressing two important drawbacks of recommender systems: sparsity and the gray-sheep problem. Both are considered based on an analysis of the power-law distribution. Data sparsity is avoided by means of inducing implicit ratings, which is affected by that type of distribution. The gray-sheep problem, which has received little attention in the area of music recommendation, is closely related to the power-law distribution, since gray-sheep users are those that listen to music that is mainly placed in the tail of the play frequency curve.

Taking into account the stated objectives, the main contributions of the work are the following:

- A procedure for inferring user-song ratings from implicit feedback in which user sessions are considered.
- A method to tackle the gray-sheep problem that involves the characterization of each user according to the play frequency of the songs he/she listen to.

An important advantage of our proposal is the fact that it requires only information about the plays of songs by users, without the need for content information. This in turn leads to a reduction in the complexity of the methods used so far to make recommendations to users with unusual tastes.

The rest of the paper is organized as follows. Section 2 contains a brief description of related works. The approach proposed to improve CF-based recommendations is detailed in section 3. Results and discussion are given in section 4 and the last section of the paper is devoted to the conclusions.

## 2. Related work

The drawbacks of the recommendation methods have been the focus of many works in literature [34]. Gray-sheep, cold-start and early-rater problems have been addressed mainly by making use of content-based approaches. The content information about musical items can be extracted from their metadata, such as title, artist, year, genre or lyrics for songs and style, country, and other demographic data for artists. Recently, social tags given to items by users are also taken as content attributes of the items, and even biographies of the artists have been used to obtain content data [35]. Besides those high-level features, low-level audio features are also exploited by means of content-based methods in many works. Spectrum, rhythm and harmony-conforming chord structure are used in [52] to determine music similarity. In [22], music is classified in melody styles as a preliminary step to learning user music preferences by mining the melody patterns from the music access behavior of the users. Pitch, tempo, loudness and entropy features are taken in [7] to classify musical items. Metadata, including title, artist, genre, and the lyrics of a musical piece, is used as content information by the recommendation process. A clustering technique is proposed in [5] to group similar songs from audio features. The aim is to provide users with recommendations from the appropriate clusters according to their listening behavior. The cold-start problems are addressed in [35] by means of deep network architectures used to combine user feedback data with artist and track embeddings. These are learned from biographies and audio signals, respectively. In [27], temporal, timbre, and rhythm features, jointly with tags provided by users, are used in a method for recommending appropriate music for videos. Each video or music is represented as a linear combination of latent factors of their associated features, and this model is used to calculate similarities on new feature spaces. Low-level description of the music is also used in [18] for emotion recognition and genre classification. These two features are learned by means of a recurrent and later used as input of a Support Vector Mashing (SVM) in order to improve its results against the use of the music original features as input.

Since content-based methods usually produce worse results than CF, hybrid approaches have been proposed for improving recommendation reliability while addressing some of the problems mentioned above [31, 50, 29]. They have also been used in the music application domain for the same purpose. In [56], unobservable user preferences are represented as a set of latent variables associated with ratings and content data, which are statistically estimated and introduced in a Bayesian network called a three-way aspect model. The hybrid proposal presented in [27] combines a content-based model for recommending unrated music, a collaboration algorithm for recommendations based on other users' suggestions and an emotion-based recommendation procedure that determines interesting music for users by computing the differences between user interests and musical emotions. A weighting system based on user listening behavior is used to combine the three methods. A questionnaire that users must fill out is necessary to discover their interests, which can be a drawback since users are not always willing to do so.

In the last years, social information is being incorporated into recommender models, either as additional attributes in CF or in hybrid recommender systems [16, 9]. In [13], topics associated with songs are induced from social tagging. Social tags assigned to songs are used in [48] to establish the similarity between them as well as to capture user preferences. Another more innovative use of social tagging is the inference of user expertise in order to find more trusted neighbors for CF [41]. Friendship relations between users of streaming platforms is a different type of social information that can be treated jointly with user preferences in order to improve music recommendations [42].

There are fewer works in the literature specifically focused on improving recommendations for the gray-sheep users, in spite of the fact that the rest of the users being affected by this problem, since it has been proven that the existence of a large number of individuals with unusual preferences might have an important impact in the recommendation quality of the entire community [10]. Content-based and hybrid methods, as previously described, can produce some improvement but are not usually very significant. Moreover, they require additional information that may not be available. Semantic Web Mining can also be used to solve gray-sheep and other typical problems of recommender systems. Semantic information is added to the existing data in order to formalize and classify product and user features. In this way, content-based models at different abstraction levels can be generated to provide recommendations based on those taxonomies. They can be combined with other approaches in order to improve recommendations [20, 34]. In [4], the authors make use of domain ontologies to classify users and items in a multi-layered community of interests prior to the similarity computation. The main drawback of this type of method is the fact that they are not easily extendible since every application domain would involve the time-consuming task of defining a specific ontology. A different approach in this line is presented in [55], where a framework for semantic-aware recommendations is proposed. In this work, concepts are automatically extracted from heterogeneous information sources, and relations between concepts are established on the basis of temporal-spatial information. The procedures involved in the framework are complex and are defined for a specific application scenario. In general terms, the methods reported in the literature to address the gray-sheep problem are very complex.

Clustering is an alternative and simpler procedure to treat users with few neighbors [11]. In [10], an extensive review of recommender systems based on diverse clustering techniques is reported. The work also includes a new proposal involving the application of the k-means algorithm to generate clusters in order to detect the gray-sheep users and a recommendation procedure for them based on their profiles. In addition, a clustering-based collaborative filtering algorithm is used to give recommendations to the remaining users. They also analyze the effect of different distance metrics in the quality of the recommendations. In some works, the clustering technique is used to address the sparsity and gray-sheep problems at the same time since some authors consider that both problems are related. In [29] and [30], fuzzy class association rules are induced from previously clustered data in order to assign more than one cluster to each user with different degrees of belonging. A simulated scenario for gray-sheep users proved the effectiveness of the method. The process, implemented in a tourist system, is not simple and requires user and items features. The Last.fm dataset is used in [47] to validate a hierarchical agglomerative clustering method for recommending resources in folksonomies, which considers the users' current navigation context in cluster selection. As far as we know, most of the methods proposed for dealing with the gray-sheep problem make use of user and/or item attributes.

The sparsity problem, caused by the insufficient number of ratings, has been widely studied in the literature. Apart from content-based methods, there are two main approaches to deal with this drawback: Matrix factorization and the use of implicit user feedback to derive ratings. Matrix factorization methods have the peculiarity that can be applied with both explicit and implicit ratings. They are procedures for dimensionality reduction that generate latent factors for each user and each item. The most extended technique in the area of recommender systems for factorizing the rating matrix is Singular Value Decomposition (SVD) [2]. In some application domains, SVD yields more reliable recommendations than standard CF algorithms [45]. However, it has a high computational cost in large scale systems; thus, less expensive SVD-based approaches, such as incremental SVD have been proposed [49]. Sparsity has also been addressed by means of other SVD-based techniques for dimensionality reduction as Latent Semantic Indexing (LSI) [14] and Principal and Component Analysis (PCA) [12].

In the music recommender area, there are some works in which matrix factorization-based procedures have been proposed. The proposal of [15] using Weighted Matrix Factorization (WMF) with implicit ratings in recommender systems has been taken in [53] as a basis of their method for song recommendation where latent factors for a given song are predicted from its audio signal. In [21], WMF is also used with the same purpose but using the number of song plays as implicit feedback.

A way to address the sparsity problem when using implicit feedback is presented by Yu et al. [57]. A model that combines the Poisson factor model and the Bayesian personalized ranking is proposed to learn user preferences and item characteristics from the frequency of interactions between users and items. Implicit ratings are also usually obtained from purchase records. In [24], log files of a mobile web application are used to identify actions, such as purchases, pre-listening and clicks, in user sessions. This information regarding the purchasing behavior of users is aimed at obtaining implicit ratings. As stated previously, those data are not available when consuming music through streaming services; thus, the usual way of obtaining implicit feedback in that context is making use of the frequency of plays. This information is provided in the Last.fm database and used in some research works where different functions for transforming it into ratings are proposed [23, 54]. However, other kinds of information can be used, such as the access history of users, which is taken in [7] to obtain user interests in a music recommendation system based on music and user groupings. In [37], a session-based collaborative filtering recommendation method is proposed, which can be used to recommend the next song the user should listen to, even when no previous user rating data is available. This method uses the items selected in the active user session to find the most similar sessions and generate the recommendation from them.

## 3. Improving CF approaches for song recommendation

The main advantage of the proposal for recommending songs presented here is the fact that only data about the plays of the songs by each user is required. Since this information is collected by the streaming systems in an easy and regular way, some drawbacks regarding the need to acquire additional data, as explicit ratings, music metadata or audio features, are avoided. The work is the continuation of a previous proposal for artist recommendation [34] and another preliminary study [40], which has been extended and adapted for recommending songs. The improvement of results compared to the main CF methods is achieved by focusing on two major aspects, a new way of obtaining implicit ratings from user sessions and the characterization of users according to the place of the songs played by them in the power-law distribution of play frequency. These approaches are ways of dealing with sparsity and gray-sheep problems, respectively. The first objective is achieved by significantly increasing the number of ratings about songs since every song played by the user will have an associated implicit evaluation. The second is addressed by characterizing each user according to their gray-sheep level.

The procedure for computing implicit ratings differs from other approaches based on frequency functions since not only the count of plays is used but also the position of the song in the user sessions. Concerning the gray-sheep treatment, there is no need for content information or the creation of clusters for different types of users, as most of the proposals in the literature do. The recommendation method is applied in the same way to all users in the system, taking into account an additional attribute that characterizes them according to the degree to which their tastes are unusual.

## 3.1 Computing implicit ratings from user sessions

Obtaining both the implicit and explicit ratings required by collaborative filtering methods always requires some type of user interaction. In the case of explicit ratings, users assign a value to items that indicates the degree to which he or she has liked that product, while implicit ratings are usually obtained from other kinds of interactions with items, such as the purchase of a product, the time spent viewing information about the item, etc. Therefore, in both cases the only available information on the preferences is about those items that have been the object of the user interaction. The aim of the recommendations is to help users discover products or services that they do not know and that they might like. Thus, only items that the user has not previously interacted with are recommended.

Traditional ways of obtaining implicit ratings for items from purchase records, clicks, or timestamp information are not possible in the context of our study since the interaction mode of users with songs in music streaming platforms is quite different from interaction with other items in other kinds of systems. Usually, binary values or simple frequency functions of plays are used to derive preferences from user implicit feedback. However, in this work, we propose a more complex model to infer users' interests from their behavior in a more reliable way.

This approach takes into account the sessions in which users play songs through the streaming services as well as a play frequency percentile function in the calculation of the ratings. Although all the songs played by the user have been chosen by him/her, the method is based on the fact that the first song in a user session is important since it has a higher probability of being a direct choice of the user at this time than the songs in other positions.

A user session is considered a period in which the user is listening to songs without interruption. It consists of songs that are played in a particular order; thus, it can be characterized as a Markov chain where initial probabilities are proportional to the number of times a given state was visited. In our case, the problem is simplified since only the start and non-start of a session is considered for the songs belonging to a session. Therefore, we use the number of times for each user that each song was at the start of the session and the number of times that each song was not at the start of a session to induce the ratings.

Let's consider a set of users $U$ and a set of songs $G$ where $u_i \epsilon U, i = 1, \ldots, n$ and $g_j \epsilon G, j = 1, \ldots, m$ represent a user and a song, respectively. In that way, in our method, the frequency function for a user $i$ and a song $j$ is computed as follows:

$$\text{Freq}_{i,j} = \alpha\, \frac{s_{i,j}}{\sum_j s_{i,j}} + (1-\alpha)\, \frac{ns_{i,j}}{\sum_j ns_{i,j}} \tag{1}$$

Where $s_{i,j}$ is the number of times the song $g_j$ was the start of the session for the user $u_i$ and $ns_{i,j}$ the number of times it was played in other positions of the sessions. The α parameter is used to adjust the importance of each term of the equation.

Once the session-based frequency is computed, Pacula's procedure is applied to obtain the ratings. This method has proven to be more suitable in the context of artist recommendation when play frequencies have a clear power-law distribution since there are few highly played artists, and most of them have few plays. The same distribution is presented when songs are the target of the recommendations, so it is also indicated in this case.

The method is based on the assumption that a user likes more a song that he/she listens to more times than one that he/she listens to less times. Therefore, rating values are given in comparative terms for each user. The implicit rating $r_{i,j}$ for the user $u_i$ and the song $g_j$ is calculated from $Freq_{i,j}$ as follows.

Let's consider that songs played by the user $u_i$ are ordered by their frequency values for this user, and $Freq_{k'}(i)$ denotes the frequency $Freq_{i,j}$ of a song $g_j$ with rank $k'$, being $k' = 1$ for the song having the highest frequency.

$$S(i) = sequence\ \{Freq_{k'}(i)\} \mid Freq_{k'}(i) > Freq_{k'+1}(i) \tag{2}$$

Then, the rating for a song with rank $k$ is computed as a linear function of the frequency percentile:

$$r_{i,j} = 4\ (1 - \textstyle\sum_{k'=1}^{k-1} Freq_{k'}(i)) \tag{3}$$

The values of the ratings are real numbers in the interval (0,4]. Unlike other item interaction-based approaches, where binary ratings are obtained based on whether or not interaction has occurred, this approach more closely resembles explicit ratings that are usually within a range of values, which can be integer or real.

The ratings calculated in that way are used in the collaborative filtering approach proposed in this work, but they also can be used in any CF method following the same procedure used when ratings are explicit.

## 3.2 User characterization-based CF approach

In order to deal with the gray-sheep problem suffered by users with uncommon tastes, we propose a procedure for characterizing users according to the play frequency of the songs they listen to. As indicated, the frequency of plays of the songs follows a power-law distribution, also called "long tail" in the context of music recommender systems. Then, gray-sheep users are those that listen to very few played songs, which are placed at the end of the long tail. However, in our proposal, it is not necessary to identify those special users, but all users in the system are associated with a gray-sheep degree along the power-law distribution curve depending on the position on the curve where the songs they play are located.

The first step of the procedure for user characterization is to determine a coefficient for the songs that reflects their popularity. This is the listening coefficient, which is computed for each song from both the number of users who play it and the number of plays it has. It is important to take into account both aspects since this coefficient will be used to characterize users, and gray-sheep ones are distinguished by having few neighbors.

For the set of users $U$ and the set of songs $G$, the number of times that user $u_i$ plays a song $g_j$ is denoted as $p_{i,j}$. This information for all users and songs is represented by the matrix of plays $\mathbf{P} := p_{i,j}$ where $\mathbf{P} \, \epsilon \, M_{n \times m}(\mathbb{N})$:

$$\boldsymbol{P} = \begin{bmatrix} p_{1,1} & \cdots & p_{1,m} \\ \vdots & \ddots & \vdots \\ p_{n,1} & \cdots & p_{n,m} \end{bmatrix} \tag{4}$$

The listening coefficient ($l_j$) for a song $g_j$ is computed as indicated in eq. 5.

$$l_j = \frac{TU_j}{\overline{TU}} \frac{\Sigma_i \left( \frac{p_{i,j}}{\overline{p_i}} \right)}{\left[ \left( \Sigma_i \Sigma_j \left( \frac{p_{i,j}}{\overline{p_i}} \right) \right) / |G| \right]} \tag{5}$$

Where $TU_j$ is the number of users who play the song $g_j$, $\overline{TU}$ is the average number of users per song and $\overline{p_i}$ the average number of plays per song of user $i$.

The coefficient $l_j$ captures the playing behaviour of the users with respect to each song. First, in the form of proportion of users who have listened to the song and second, in the form of number of plays of the song by a given user with respect to the average number of plays of this user. A normalized listening coefficient $L_j$ can be obtained by means of eq. 6.

$$L_{g_j} = \frac{l_j - \min l_j}{\max l_j - \min l_j} \Rightarrow L_j \in [0,1] \tag{6}$$

In the next step, the User Playing Coefficient (UPC) that characterizes users is computed from the listening coefficients of the songs they listen to (eq. 7).

$$UPC_i = \frac{\Sigma_j \beta_{i,j} L_j}{TG_i} \tag{7}$$

Being $\beta_{i,j}$ a parameter that takes the value 1 if the song $g_j$ has been played by the user $u_i$ and the value 0 otherwise. $TG_i$ is the total number of songs played by user $u_i$. Users with high values of UPC have preferences in common with many others, while those with low values would be gray-sheep users.

Both user-playing coefficients and session-based implicit rating are needed for the next step that involves the CF method proposed in this work and described in the next subsection. Algorithm 1 describes the complete sequence of steps required for their calculation.

Algorithm 1. Computing session-based implicit ratings and User Playing Coefficients (UPC)

1: $\mathbf{P} := p_{i,j}, \mathbf{P} \in M_{n \times m}(\mathbb{N})$
2: $\mathbf{S} := s_{i,j}, \mathbf{S} \in M_{n \times m}(\mathbb{N})$
3: $\mathbf{NS} := ns_{i,j}, \mathbf{NS} \in M_{n \times m}(\mathbb{N})$
4: Set $\alpha$ value
5: **for** $i = 1$ *to* $n$ **do**
6:     **for** $j = 1$ *to* $m$ **do**
7:       $Freq_{i,j} = \alpha \left(s_{i,j}/\sum_{j=1}^{m} s_{i,j}\right) + (1-\alpha)(ns_{i,j}/\sum_{j=1}^{m} ns_{i,j})$
8:     **end for**
9: **end for**
10: **for** $i = 1$ *to* $n$ **do**
11:     $T = sequence\left\{(Freq_{i,j})_k\right\} \forall j \mid p_{i,j} > 0 \wedge (Freq_{i,j})_{k'} > (Freq_{i,j})_{k'+1}$
12:     $S(i) = sequence\left\{Freq_{k'}(i)\right\} = T$
13:     **for** $j = 1$ *to* $m$ **do**
14:       Set $k$ value $\mid Freq_k(i) = Freq_{i,j}$
15:       $r_{i,j} = 4\left(1 - \sum_{k'=1}^{k-1} Freq_{k'}(i)\right)$
16:     **end for**
17: **end for**
18: $\mathbf{R} := r_{i,j}, \mathbf{R} \in M_{n \times m}(\mathbb{N})$
19: **for** $j = 1$ *to* $m$ **do**
20:     $TU_j = |U_j|, U_j \subseteq U \mid u_i \in U_j \forall i \mid p_{i,j} > 0$
21:     $TP_j = \sum_i p_{i,j}$
22: **end for**
23: $\overline{TU} = \sum_{j=1}^{m} TU_j/m$
24: $\beta_{i,j} = 0 \forall i,j$
25: **for** $i = 1$ *to* $n$ **do**
26:     $TG_i = |G_i|, G_i \subseteq G \mid g_j \in G_i \forall j \mid p_{i,j} > 0$
27:     $\overline{p_i} = \sum_{i=1}^{n} p_{i,j}/TG_i$
28:     $\beta_{i,j} = 1 \forall j \mid p_{i,j} > 0$
29: **end for**
30: **for** $j = 1$ *to* $m$ **do**
31:     $l_j = (TU_j/\overline{TU}) \left(\sum_{i=1}^{n}(p_{i,j}/\overline{p_i})/(\sum_{i=1}^{n}\sum_{j=1}^{m}(p_{i,j}/\overline{p_i}))/m\right)$
32:     $L_j = (l_j - min\ l_j)/(max\ l_j - min\ l_j)$
33: **end for**
34: **for** $i = 1$ *to* $n$ **do**
35:     $UPC_i = \sum_{j=1}^{m} \beta_{i,j} L_{g_j}/TG_i$
36: **end for**
37: $\mathbf{UPC} := UPC_i, \mathbf{UPC} \in M_{n \times 1}(\mathbb{N})$
38: **Output**: R and UPC

### 3.3 Incorporating UPC to user-based CF

In user-based collaborative filtering, active users receive recommendations of items liked by their nearest neighbors. Two users are defined as neighbors if they have some items in common that they have rated with close scores. In the context of our work, users who like the same songs would have similar ratings and would, therefore, be neighbors.

For the set U of $m$ users and the set $G$ of $n$ songs, there is a list of ratings for each user $u_i$ that user has given to a subset of songs $G_{ui}$, where $G_{ui} \subseteq G$. Ratings are stored in a $m \times n$ matrix called the rating matrix, where each element is the rating that a user $u_i$ gives to a song $g_j$.

$$\mathbf{R} := r_{i,j} \,,\ \mathbf{R} \, \epsilon \, M_{n \times m}(\mathbb{N}) \qquad (8)$$

When explicit ratings are used, this matrix usually has many null elements, because users have rated a small subset of songs, in a way that the fewer the number of rated items, the sparser the matrix. As stated, this is an important problem inherent to CF methods that can be minimized by making use of implicit feedback. In this work, the rating matrix contains the session-based implicit ratings computed by means of eq. 1 to 3, as described in the algorithm 1. Our proposal also requires the User Playing Coefficients for every user ($UPC_{u_i}$), whose computing procedure is included in the same algorithm.

In order to make recommendations to the active user $u_a$ it is necessary to find user neighbors. Among the metrics that can be applied to computer user similarity, the Pearson correlation coefficient and cosine similarity are the most frequently used in the field of recommender systems.

The Pearson correlation coefficient evaluates the linear relationship between two variables and is obtained from its covariance. This coefficient $\omega\,(u_a, u_i\,)$ for the active user $u_a$ and another user $u_i$ is computed as follows.

$$\omega\,(u_a, u_i\,) = \frac{\Sigma_j(r_{aj} - \bar{r}_a)(r_{ij} - \bar{r}_i)}{\sqrt{\Sigma_j(r_{aj} - \bar{r}_a)^2(r_{ij} - \bar{r}_i)^2}} \qquad (9)$$

Where $r_{aj}$ and $r_{ij}$ are the ratings of user $u_a$ and user $u_i$ for song $g_j$ respectively, and $\bar{r}_a$ and $\bar{r}_i$ are the average ratings of user $u_a$ and user $u_i$ respectively. The Pearson coefficient can represent inverse and direct correlation with its values in the interval [-1, 1], where the value 0 corresponds to the absence of correlation.

Another commonly used similarity metric is cosine, which is given by the dot product of the vectors representing the preferences of two given users, $u_a$ and $u_i$ in the Euclidean space. The cosine similarity between those users is computed according to eq. 10, where $V_{u_a}$ and $V_{u_i}$ are the vectors containing the implicit ratings for songs corresponding to users $u_a$ and $u_i$ respectively.

$$cosSim\,(u_a, u_i\,) = cos\left(V_{u_a}, V_{u_i}\,\right) = \frac{V_{u_a} \cdot V_{u_i}}{\|V_{u_a}\| \, \|V_{u_i}\|} \qquad (10)$$

This is the metric used in our approach since it can be used to compute similarity from other user attributes in addition to ratings. The additional attribute incorporated at this point is the User Playing Coefficient, $UPC_{u_i}$, which influences the search result of the k-nearest neighbors. To do this, an attribute-aware weighted user-based K-NN approach was applied [51]. Specifically, the implementation provided by the Recommender extension of RapidMiner [32, 33]. The resulting weighted similarity ($wCosSim$) between a user $i$ and the active user, along with their ratings, are used to predict the rating that a given user would assign to a song $g_j$ that he/she has not played yet, by means of eq. 11 [1]. Only the $k$ nearest neighbors, that is, those with the highest similarity values, will be taken into account to make the predictions.

$$pr_{aj} = \bar{r}_a + \frac{\sum_{i=1}^{k} wCosSim(u_a, u_i)(r_{ij} - \bar{r}_i)}{\sum_{i=1}^{k} |sim(u_a, u_i)|} \tag{11}$$

The results of applying this proposal have been compared to those provided by other CF methods. They are analyzed in the next section.

## 4. Comparative evaluation of the proposal versus other CF methods

In order to validate the recommendation approach, a comparative study was conducted in which this proposal and other widely used CF methods were applied to a dataset containing real data collected by Oscar Celma (https://www.upf.edu/web/mtg/lastfm360k) from the Last.fm streaming platform. Only information concerning the play of songs by users was used in the study. Specifically, 420,209 records corresponding to 86,000 songs played by 53 users over two years were processed. Each of them consisted of the user ID, the song and timestamp when the song was played.

The first step of the preprocessing process was to establish user sessions and place the songs played in each session in the order in which they were listened to. We considered a user inactivity period longer than 15 minutes as the mark of the end of a session. Then, the first play after that period was the indication of the start of a new session. After determining the sessions, the second step was to compute implicit ratings from the count of plays in each session and the number of times that each song was the start of the session for each user.

In order to compare our proposal based on sessions to the classical implicit ratings calculation based on the frequency of plays, we applied both methods. For the first, we reported the results obtained for two values of the alpha parameter. In the second, Pacula's method was used to calculate the ratings from the simple play count, without considering sessions. Basic and matrix factorization CF methods were tested to check whether the new rating computation procedure succeeded in increasing the reliability of the recommendations. One of these methods was K-nearest neighbor (K-NN) which is extensively used in the implementation of recommender systems. We tested user-based K-NN using both cosine and Pearson similarity measures for determining the neighborhood of K users who have preferences most similar to those of the active user. The number of K neighbors was set to 5 since it provided the best results in the experiments. Although it may seem insufficient to make the predictions, this number of neighbors has been successfully used in other work in the same field of application [43]. In addition, two matrix factorization methods were applied, the basic technique and a variant called biased matrix factorization that incorporates user and item regularization parameters. Ten-fold cross-validation was performed to evaluate the results, and the metrics used were RMSE (Root-Mean-Square Error), MAE (Mean Absolute Error) and NMAE (Normalized Mean Absolute Error).

Figure 1 shows the error rates of the K-NN output for the cosine distance and Pearson coefficient. The results of matrix factorization methods are shown in Figure 2. We can see that the error rates decrease for all the methods when the session-based approach is used, especially with $\alpha = 0.7$. This difference is significantly greater in the case of matrix factorization methods, which yielded worse results than k-NN. While the NMAE reduction achieved with session-based ratings ($\alpha = 0.7$) versus play count-based rating is 17.12% and 16.08% for K-NN with cosine similarity and Pearson coefficient, respectively, the reduction versus matrix factorization (MF) and biased matrix factorization (BMF) is 37.56% and 24.43%, respectively.

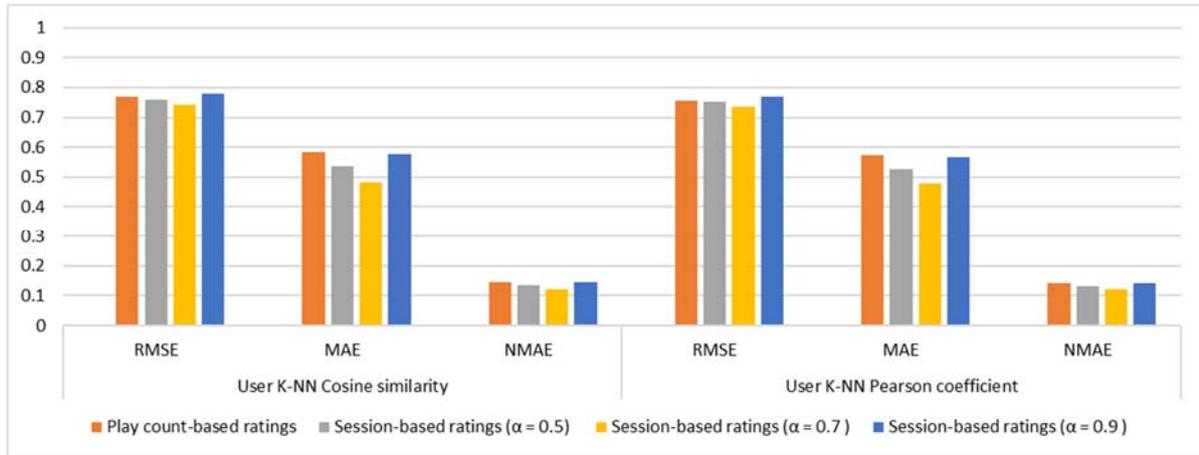

Figure. 1: Error rates given by user K-NN algorithm with two similarity metrics for the ratings computed in four different ways.

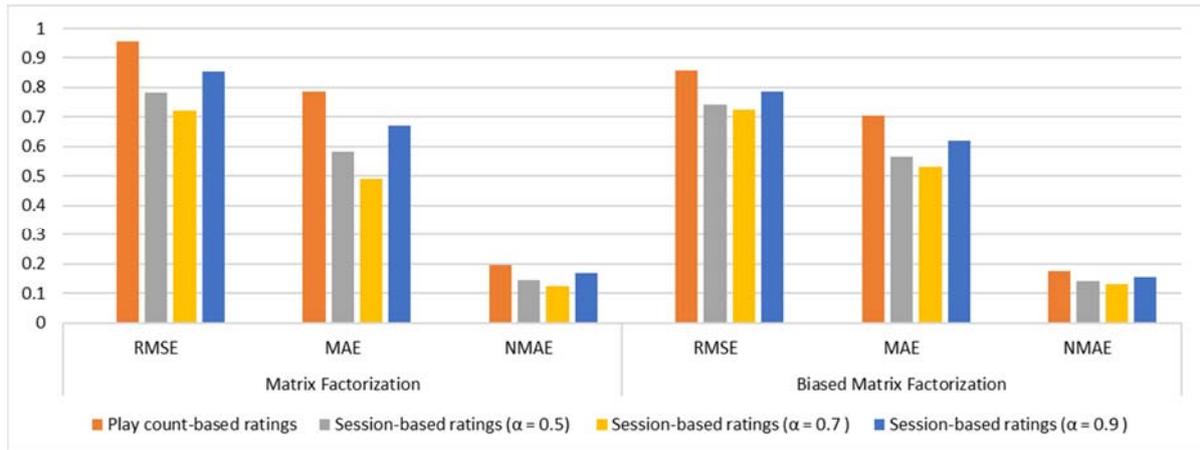

Figure. 2: Error rates given by two matrix-factorization methods for the ratings computed in four different ways.

Once the method for obtaining the ratings was validated, we checked the impact of introducing user characterization in CF. User playing coefficients, which characterize the degree in which a specific user is a gray-sheep, were computed according to the procedure described in subsection 3.2. These coefficients ($UPC_{u_i}$) were discretized before applying the proposed CF method aimed at improving efficiency. The results for different numbers of bins were analyzed to obtain the optimal partition. As expected, the errors decreased as the number of intervals increased, tending to stabilize at 300 bins. Thus, this was the number of bins chosen to conduct the experiments. Discretized coefficients $UPC_{u_i}$ obtained for every user $u_i$, were used in the user-based K-NN algorithm as an additional attribute to compute user similarity, making use of the cosine measure, as described in subsection 3.3. This approach that we call "User attribute K-NN UPC" was also tested with ratings based on play count and ratings based on sessions for three different values of $\alpha$ (eq.1-3) and its results compared to those provided by the CF methods tested above. Table 1 shows the detailed results of all these methods.

Table 1: Error rates of the methods involved in the study

| Rating | User attribute K-NN UPC | | |
| | *RMSE* | *MAE* | *NMAE* |
|---|---|---|---|
| Play count-based | 0.701±0.007 | 0.539±0.004 | 0.135±0.001 |
| Session-based (α = 0.5) | 0.693±0.009 | 0.497±0.006 | 0.124±0.001 |
| Session-based (α = 0.7) | 0.691±0.010 | 0.457±0.006 | 0.114±0.002 |
| Session-based (α = 0.9) | 0.707±0.007 | 0.533±0.005 | 0.133±0.001 |
| | **User K-NN Cosine distance** | | |
| | *RMSE* | *MAE* | *NMAE* |
| Play count-based | 0.771± 0.007 | 0.583±0.004 | 0.146±0.001 |
| Session-based (α = 0.5) | 0.760±0.009 | 0.535±0.006 | 0.134±0.001 |
| Session-based (α = 0.7) | 0.743± 0.011 | 0.484±0.006 | 0.121±0.002 |
| Session-based (α = 0.9) | 0.778±0.008 | 0.576±0.005 | 0.144±0.002 |
| | **User K-NN Pearson coefficient** | | |
| | *RMSE* | *MAE* | *NMAE* |
| Play count-based | 0.756±0.007 | 0.572±0.004 | 0.143±0.001 |
| Session-based (α = 0.5) | 0.753±0.009 | 0.528±0.006 | 0.132±0.001 |
| Session-based (α = 0.7) | 0.737±0.012 | 0.479±0.006 | 0.120±0.002 |
| Session-based (α = 0.9) | 0.768±0.007 | 0.567±0.005 | 0.142±0.002 |
| | **Matrix factorization** | | |
| | *RMSE* | *MAE* | *NMAE* |
| Play count-based | 0.955±0.007 | 0.787±0.006 | 0.197±0.002 |
| Session-based (α = 0.5) | 0.784±0.009 | 0.583±0.006 | 0.146±0.002 |
| Session-based (α = 0.7) | 0.723±0.012 | 0.492±0.006 | 0.123±0.002 |
| Session-based (α = 0.9) | 0.854±0.009 | 0.671±0.006 | 0.168±0.002 |
| | **Biased matrix factorization** | | |
| | *RMSE* | *MAE* | *NMAE* |
| Play count-based | 0.859±0.004 | 0.704±0.004 | 0.176±0.001 |
| Session-based (α = 0.5) | 0.742±0.009 | 0.564±0.006 | 0.141±0.001 |
| Session-based (α = 0.7) | 0.724±0.011 | 0.533±0.006 | 0.133±0.002 |
| Session-based (α = 0.9) | 0.787±0.009 | 0.619±0.006 | 0.155±0.001 |

One of the main conclusions obtained from the table is the confirmation that the session-based rating with $\alpha = 0.7$ provides the best results. Another observation regarding the metric values in the table, is the more significant error reduction for MAE than for RMSE. It is known that MAE is a linear score that is not as sensitive to outliers as RMSE, which further penalizes large errors. Therefore, the smaller decreasing of RMSE values may be due to the fact that there could be some predictions where the deviation from the actual value is significantly higher than the majority, both when using our proposal and the other methods. That can mask the improvement of the rest of the predictions.

From the analysis of the table, we can also derive that the lowest error rates occur with the new UPC-based method, regardless of the type of rating used. Figures 3, 4 and 5 representing RMSE, MAE and NMAE respectively allow us to visualize jointly both facts. It can be seen that the line representing the User attribute K-NN UPC method is in the lowest position, and the lowest points of all the lines representing the methods are those corresponding to the session-based ratings with $\alpha = 0.7$.

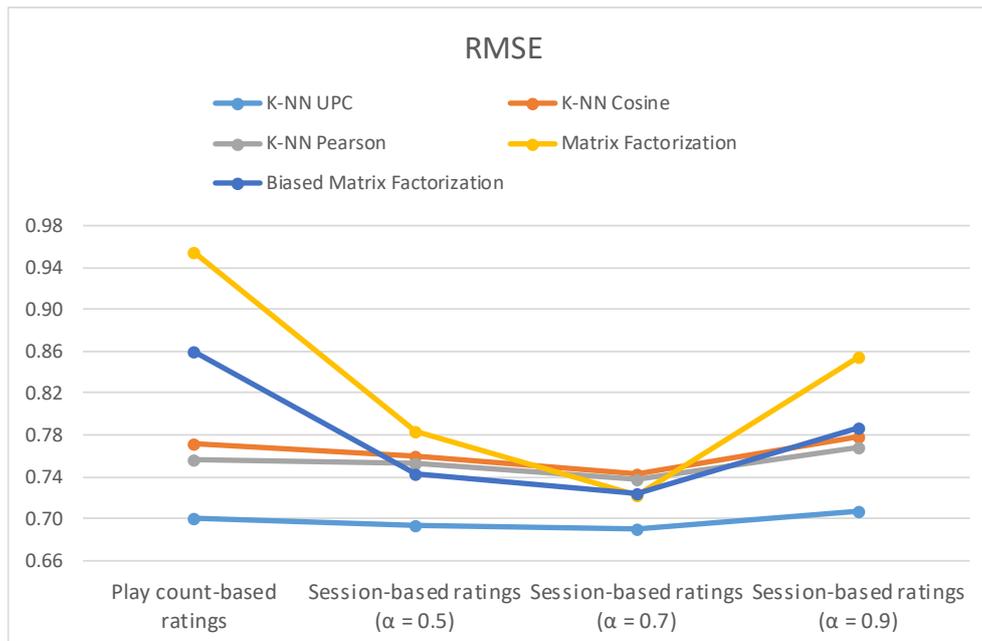

Figure. 3: RMSE of the CF methods tested in the study for ratings calculated in four different ways.

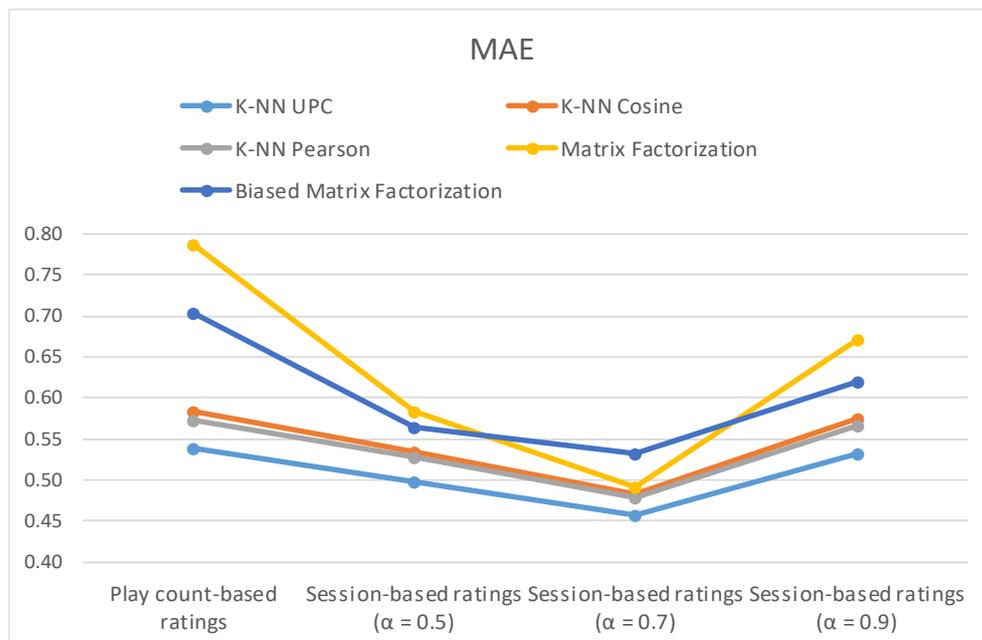

Figure. 4: MAE of the CF methods tested in the study for ratings calculated in four different ways.

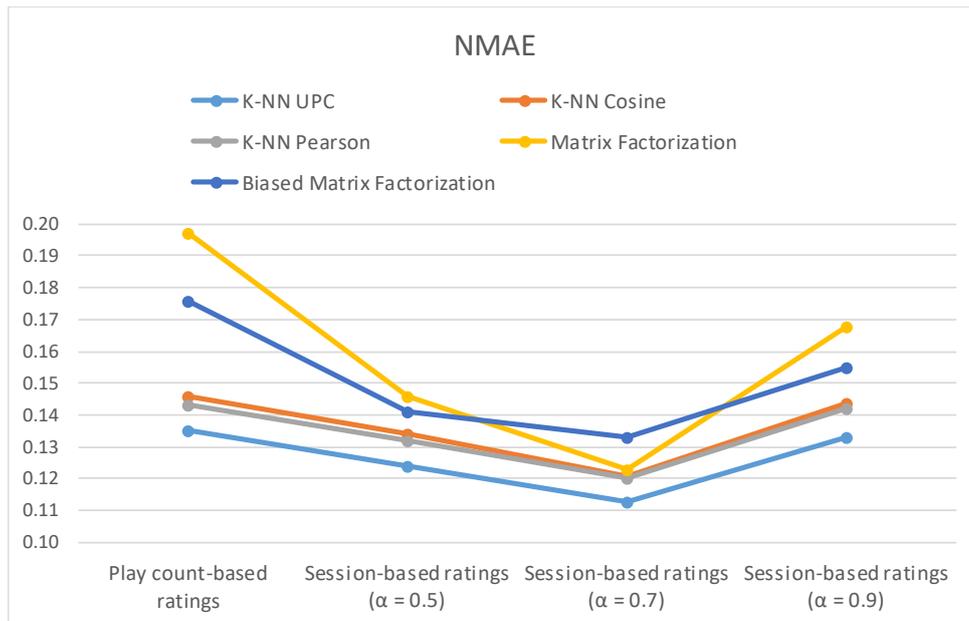

Figure. 5: NMAE of the CF methods tested in the study for ratings calculated in four different ways.

In order to get an idea of the improvement achieved with the proposed approaches, we can compare the NMAE result of User attribute K-NN UPC with session-based ratings ($\alpha = 0.7$) to both the best and the worst NMAE result of the other CF methods using classical play count-based rating. NMAE was reduced by 20.28% with respect to the best, User K-NN with Pearson coefficient, and by 42,64% with respect to the worst, Matrix Factorization. Differences are also important, comparing the UPC-based method to the rest of the methods when using session-based ratings ($\alpha = 0.7$) in all of them. Figure 6 shows these differences. In this case, the improvements provided by user attribute K-NN UPC vary between 5.00% and 14.29%.

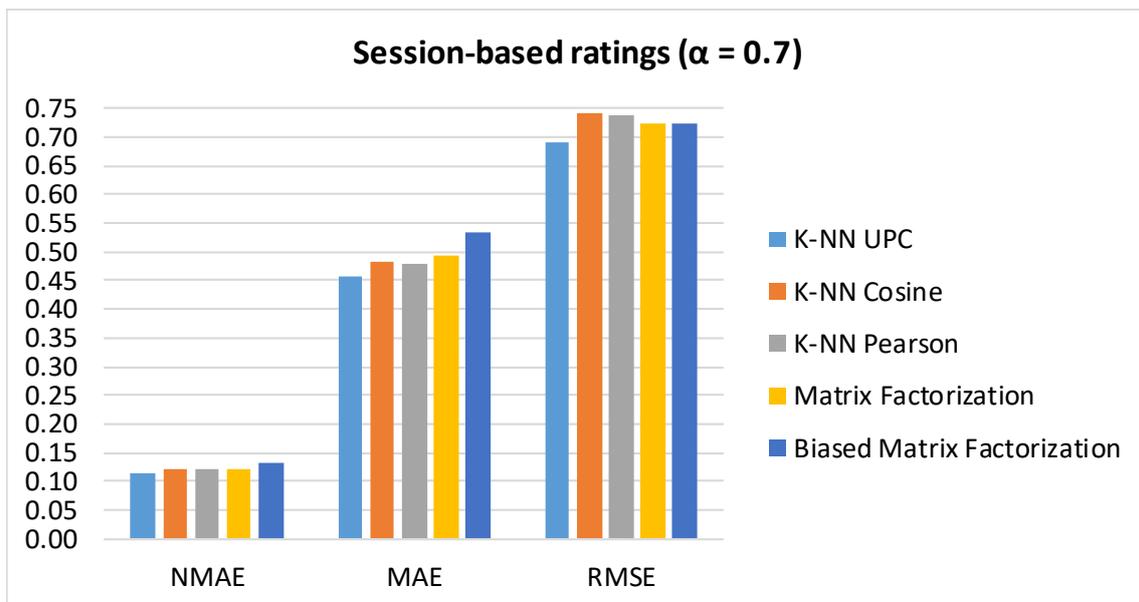

Figure. 6: Comparison of the methods tested in the study with session-based ratings ($\alpha = 0.7$).

# Conclusions

Sparsity and gray-sheep problems are two of the main reasons CF methods do not provide the reliability required in some recommendation systems. Both have been addressed in many works in the literature, although in the field of music, they have been less studied, especially the second. This is a major drawback because some of the proposed solutions are difficult to implement in this application domain. On the one hand, the way of obtaining implicit feedback in streaming services is totally different from other web applications due, among other reasons, to the fact that there are no individualized purchase records of songs, and the mode of consuming music is different from the consumption of other items. On the other hand, most of the proposals to deal with the problems of sparsity and gray-sheep, particularly with the latter, make use of content information that is difficult to obtain and that, in many cases, does not lead to the expected results.

In this work, an approach to improve recommendation reliability in the context of music streaming services is presented. Its main value is to address implicit rating computation and user characterization only from the play timestamp of the songs, information that is regularly collected by streaming platforms. The procedure proposed in this work for obtaining the ratings differs from most of the methods, that generally use play counts, because this procedure is based on user sessions. Furthermore, a new way of managing gray-sheep users based on the long tail distribution is presented. The results show a significant improvement of recommendation reliability over traditional CF and matrix factorization methods.

## Data Availability

The dataset used in the study is publicly available for non-commercial use at
https://www.upf.edu/web/mtg/lastfm360k

## Conflicts of Interest

The authors declare that there is no conflict of interest regarding the publication of this paper.

## Funding Statement

This research has been supported by the Department of Education of the Junta de Castilla y León, Spain (ORDEN EDU/667/2019), [SA064G19].

## References

[1]     Aggarwal, C.C. (2016). Recommender Systems, Springer.

[2]     Billsus, D. & Pazzani, M. (1998). Learning collaborative information filters. In Proceedings of the 15th International Conference on Machine Learning (ICML '98) (pp. 46-54). Morgan Kaufmann, San Francisco, CA.

[3]     Breese, J. S., Heckerman, D., & Kadie C. (1998). Empirical analysis of predictive algorithms for collaborative filtering. In Proceedings of the Fourteenth Conference on Uncertainty in Artificial Intelligence (pp. 43-52). Morgan Kaufmann Publishers Inc. San Francisco, CA, USA,.

[4]     Cantador, I., Bellogín, A., & Castells, P. (2008). A multilayer ontology-based hybrid recommendation model. AI Communications, 21 (2008), 203-210.

[5]     Cataltepe, Z., & Altinel, B. (2007). Music recommendation based on adaptive feature and user grouping. In Proceedings of 22nd International Symposium on Computer and Information Sciences (pp. 1-6). Ankara, Turkey.

[6]     Celma, O. (2010). Music Recommendation and Discovery: The Long Tail, Long Fail, and Long Play in the Digital Music Space. Berlín Heidelberg, Germany: Springer.

[7]     Chen, H. C., & Chen, A. L. P. (2005). A music recommendation system based on music and user grouping. Intelligent Information Systems, 24, 113-132.

[8]     Claypool, M., Gokhale, A., Mir, T., Murnikov, P., Netes, D., & Sartin, M. (1999). Combining content-based and collaborative filters in an online newspaper In Proceedings of ACM SIGIR workshop on recommender systems, Berkeley, CA, USA: ACM.

[9]     Deng, S., Wang, D., Li, X., & Xu, G. (2015). Exploring user emotion in microblogs for music recommendation. Expert Systems with Applications, 42 (2015), 9284-9293.

[10]    Ghazanfar, M. A., & Prügel-Bennett, A. (2014). Leveraging clustering approaches to solve the gray-sheep users problem in recommender systems. Expert Systems with Applications, 41 (2014), 3261-3275.

[11]    Ghorbani, H., & Novin, A. H. (2016). An introduction on separating gray-sheep users in personalized recommender systems using clustering solution. International Journal of Computer Science and Software Engineering, 5 (2016), 14-18.

[12]    Goldberg, K., Roeder, T., Gupta, T., & Perkins, C. (2001). Eigentaste: a constant time collaborative filtering algorithm. Information Retrieval, 4(2001), 133–151.

[13]    Hariri, N., Mobasher, B., & Burke, R. (2012). Context-aware music recommendation based on latent topic sequential patterns. In Proceedings of the sixth ACM conference on Recommender systems (pp. 131-138). Dublin, Ireland.

[14]    Hofmann, T. (2004). Latent semantic models for collaborative filtering. ACM Transactions on Information Systems, 22 (2004), 89–115.

[15]    Hu, Y., Koren, Y., & Volinsky, C. (2008). Collaborative filtering for implicit feedback datasets. In Proceedings of Eighth IEEE International Conference on Data Mining (pp. 263–272). IEEE.

[16]    Hyung, Z., Lee, K., & Lee, K. (2014). Music recommendation using text analysis on song requests to radio stations. Expert Systems with Applications, 41 (2014), 2608-2618.

[17]    Iqbal, M. (2019). Spotify Usage and Revenue Statistics, https://www.businessofapps.com/data/spotify-statistics/#3. Last visited on March 20, 2020.

[18]    Jakubik, J. & Kwaśnicka, H. (2018), Similarity-Based Summarization of Music Files for Support Vector Machines, Complexity, Volume 2018, Article ID 1935938, 10 pages.

[19]    Kamehkhosh, I., Bonnin, G. & Jannach, D. (2019). Effects of recommendations on the playlist creation behavior of users. User Modeling and User-Adapted Interaction (2019). https://doi.org/10.1007/s11257-019-09237-4.

[20]    Kim, H. N., Alkhaldi, A., El Saddik, A., & Jo, G. S. (2011). Collaborative user modeling with user-generated tags for social recommender systems. Expert Systems with Applications, 38 (2011), 8488-8496.

[21]    Koh, C., & Ko. T. (2013). Deep Content-based Music Recommendation. Advances in Neural Information Processing Systems, 26. Presented at the Neural Information Processing Systems Conference (NIPS 2013), Lake Tahoe, NV, USA: Neural Information Processing Systems Foundation (NIPS).

[22]    Kuo, F. F., & Shan, M. K. (2002). A personalized music filtering system based on melody style classification. In Proceedings of the IEEE international conference on data mining (pp. 649-652). Maebashi City, Japan.

[23]    Lee, K., & Lee, K. (2015). Escaping your comfort zone: A graph-based recommender system for finding novel recommendations among relevant items. Expert Systems with Applications, 42 (2015), 4851-4858.

[24]    Lee, S. K., Cho, Y. H., & Kim, S. H. (2010). Collaborative filtering with ordinal scale-based implicit ratings for mobile music recommendations. Information Sciences, 180 (2010), 2142-2155.

[25]    Linden, G., Smith, B., & York, J. (2003). Amazon.com recommendations: Item to item collaborative filtering. IEEE Internet Computing, 7(2003), 76-80.

[26]    Liu, C. L., & Chen, Y. C. (2018) Background music recommendation based on latent factors and moods. Knowledge-Based Systems, 159 (2018), 158-170.

[27]    Lu, C. C., & Tseng, V. S. (2009). A novel method for personalized music recommendation. Expert Systems with Applications, 36 (2009), 10035-10044.

[28]    Lucas, J. P., Laurent, A., Moreno, M. N., & Teisseire, M. (2012). A fuzzy associative classification approach for recommender systems. International Journal of Uncertainty, Fuzziness and Knowledge-Based Systems, 20 (2012), 579-617.

[29]    Lucas, J. P., Luz, N., Moreno, M. N., Anacleto, R., Almeida, A., & Martins, C. (2013). A hybrid recommendation approach for a tourism system. Expert Systems with Applications, 40 (2013), 3532-3550.

[30]    Lucas, J. P., Segrera, S., & Moreno, M. N. (2012). Making use of associative classifiers in order to alleviate typical drawbacks in recommender systems. Expert Systems with Applications, 39 (2012), 1273-1283.

[31]    Melville, P., Mooney, R. J., & Nagarajan, R. (2002). Content boosted collaborative filtering for improved recommendations. In Proceedings of the 18th National Conference on Artificial Intelligence (pp.187-192). Edmonton, Canada.

[32]     Mihelčić, M., Antulov-Fantulin, N., Bošnjak, M. & Šmuc, T. (2012). Extending RapidMiner with recommender systems algorithms. In Proceedings of the 3rd RapidMiner Community Meeting and Conference (RCOMM 2012) (pp. 63-74). Shaker Verlag.

[33]    Mihelčić, M., Bošnjak, M., Antulov-Fantulin, N & Šmuc, T. (2014). Constructing Recommender Systems in RapidMiner. In Hofmann, M. and Klinkenberg, R. (eds.), RapidMiner. Data Mining Use Cases and Business Analytics Applications (pp. 119-143). CRC Press.

[34]    Moreno, M. N., Segrera, S., López, V. F., & Muñoz, M. D. (2016). Web Mining based Framework for solving usual problems in recommender systems. A case study for movies' recommendation. Neurocomputing, 176 (2016), 72-80.

[35]    Oramas, S., Nieto, O., Sordo, M., & Serra, X. (2017). A deep multimodal approach for cold-start music recommendation. In Proceedings of the 2nd Workshop on Deep Learning for Recommender Systems (pp. 32–37). New York, NY, USA: ACM.

[36]    Pacula, M.  A matrix factorization algorithm for music recommendation using implicit user feedback. http://www.mpacula.com/publications/lastfm.pdf. Last accessed March 20, 2020.

[37]    Park, S. E., Lee, S., & Lee, S. (2011). Session-based collaborative filtering for predicting the next song. In Proceedings of first ACIS/JNU International Conference on Computers, Networks, Systems and Industrial Engineering (pp. 353-358). Jeju Island, Korea.

[38]    Resnik, P. (1999). Semantic similarity in a taxonomy: An information based measure and its application to problems of ambiguity in natural language. Journal of Artificial Intelligence, 11, 94-130.

[39]    Sánchez-Moreno, D., Gil, A. B., Muñoz, M. D., López, V. F., & Moreno, M. N. (2016). A collaborative filtering method for music recommendation using playing coefficients for artists and users. Expert Systems with Applications, 66 (2016), 234-244.

[40]    Sánchez-Moreno, D., Gil, A. B., Muñoz, M. D., López, V. F., & Moreno, M. N. (2017). Recommendation of songs in music streaming services. Dealing with sparsity and gray sheep problems. In F. de la Prieta, Z. Vale, L. Antunes, T.Pinto, A. T. Campbell, V.Julian, A. Neves, & M.N. Moreno (Eds.), Trends in Cyber-Physical Multi-Agent Systems. The PAAMS Collection - 15th International Conference, PAAMS 2017. Advances in Intelligent Systems and Computing series, 619 (pp. 206-213). Springer.

[41]    Sánchez-Moreno, D., Moreno-García, M. N., Sonboli, N., Mobasher, B., & Burke, R (2018). Inferring user expertise from social tagging in music recommender systems for streaming services, In F.de Cos et al. (eds), Hybrid Artificial Intelligence Systems, Lecture Notes in Artificial Intelligence, (pp. 39-49). Springer.

[42]    Sánchez-Moreno, D., Pérez-Marcos, J., Gil, A. B., López, V. F., & Moreno-García, M. N. (2019). Social influence-based similarity measures for user-user collaborative filtering applied to music recommendation, In Rodríguez S. et al. (eds), Distributed Computing and Artificial Intelligence, Special Sessions, 15th International Conference, 2018. Advances in Intelligent Systems and Computing, 801 (pp. 1-8). Springer.

[43]    Sánchez-Moreno, D., Zheng, Y. & Moreno-García M.N. (2018). Incorporating Time Dynamics and Implicit Feedback into Music Recommender Systems. In Proceedings of the 17th IEEE/WIC/ACM International Conference on Web Intelligence (WI'18), pp. 580-585, Santiago, Chile.

[44]    Sarwar, B. M., Karypis, G., Konstan, J. A., & Riedl, J. (2002). Incremental Singular Value Decomposition Algorithms for Highly Scalable Recommender Systems. In Proceedings of 5th International Conference in Computers and Information Technology. Rousse, Bulgaria: ACM.

[45]    Sarwar, B. M., Karypis, G., Konstan, J.A., & Riedl, J. (2000). Application of Dimensionality Reduction in Recommender System-A Case Study. In Proceedings of ACM Web-mining for Ecommerce Workshop. Boston, MA, USA.

[46]    Sarwar, B., Karypis, G., Konstan, J. A., & Riedl, J. (2001). Item-based collaborative filtering recommendation algorithm. In Proceedings of the tenth International World Wide Web Conference (pp. 285-295). Hong Kong, Hong Kong: ACM.

[47]    Shepitsen, A., Gemmell, J., Mobasher, B., & Burke, R. (2008). Personalized recommendation in social tagging systems using hierarchical clustering. In Proceedings of the 2008 ACM conference on Recommender systems (pp. 259–266). New York, NY, USA: ACM.

[48]    Su, J. H., Chang, W. Y., & Tseng V. S. (2013). Personalized music recommendation by mining social media tags. Procedia Computer Science, 22 (2013), 303-312.

[49]    Su, X., & Khoshgoftaar, T. M. (2009). A survey of collaborative filtering techniques. Advances in Artificial Intelligence, 2009, 1-19.

[50]    Su, X., Greiner, R., Khoshgoftaar, T. M., & Zhu, X. (2007). Hybrid collaborative filtering algorithms using a mixture of experts. In Proceedings of the IEEE/WIC/ACM International Conference on Web Intelligence (pp. 645-649). Silicon Valley, Calif, USA.

[51]    Tso, K. & Schmidt-Thieme, L. (2006) Attribute-aware Collaborative Filtering. In: Spiliopoulou M., Kruse R., Borgelt C., Nürnberger A., Gaul W. (eds) From Data and Information Analysis to Knowledge Engineering. Studies in Classification, Data Analysis, and Knowledge Organization. Springer, Berlin, Heidelberg.

[52]    Tzanetakis, G. (2002). Musical Genre Classification of Audio Signals. IEEE Transactions on Speech and Audio Processing, 10 (2002), 293-302.

[53]    Van den Oord, A., Dieleman, S., & Schrauwen, B. (2013), Deep content-based music recommendation. Advances in Neural Information Processing Systems, 26 (2013), 2643–2651.

[54]    Vargas, S., & Castells, P. (2011). Rank and relevance in novelty and diversity metrics for recommender systems. In Proceedings of the fifth ACM conference on recommender systems (pp. 109-116). New York, NY, USA: ACM.

[55]    Wang, F., Hu, L., Sun, R., Hu, J., & Zhao, K. (2018). SRMCS: A semantic-aware recommendation framework for mobile crowd sensing, Information Sciences, 433-434 (2018), 333-345.

[56]    Yoshii, K., Goto, M., Komatani, K., & Ogata, T., Okuno, H.G. (2006). Hybrid collaborative and content-based music recommendation using probabilistic model with latent user preferences. In Proceedings of the 7th international conference on music information retrieval (pp. 296-301). Victoria, Canada: University of Victoria.

[57]    Yu, Y., Zhang, L., Wang, C., Gao, R., Zhao, W., & Jiang J. (2019). Neural Personalized Ranking via Poisson Factor Model for Item Recommendation. Complexity, Volume 2019, Article ID 3563674, 16 pages.